\documentclass[journal]{new-aiaa}
\usepackage[utf8]{inputenc}
\usepackage{textcomp}
\usepackage{colortbl}
\usepackage{xcolor}
\usepackage{graphicx}
\usepackage{amsmath}
\usepackage[version=4]{mhchem}
\usepackage{siunitx}
\usepackage{longtable,tabularx}
\usepackage{booktabs}
\definecolor{trimgreen}{HTML}{C0DD97}
\definecolor{trimred}{HTML}{F09595}

\usepackage{new-mysymb}

\title{Trajectory Optimization of Morphing Aerial Vehicles Based on Mid-Fidelity Aeroservoelastic Models}

\author{Subarna Pudasaini\footnote{Graduate Student, Department of Aerospace Engineering, Student Member AIAA}, Parker Smith\footnote{Undergraduate Student, Department of Aerospace Engineering, Student Member AIAA}, Daning Huang\footnote{Assistant Professor, Department of Aerospace Engineering, Senior Member AIAA}}
\affil{Pennsylvania State University, University Park, PA, 16801}

\begin{document}

\maketitle

\begin{abstract}
Morphing aerial vehicles offer enhanced maneuverability and fuel efficiency compared to fixed-wing configurations. However, the trade-off between performance gains and control cost in dynamic, unsteady maneuvers remains under-explored. This paper addresses this by integrating a trajectory optimization framework with a mid-fidelity aeroservoelastic model, coupling nonlinear multi-body structural dynamics with an unsteady vortex lattice method. A physics-based control cost model captures the energy required to overcome instantaneous aerodynamic hinge moments. Applied to an aircraft with flexible, high-aspect-ratio wings and morphing winglets, the framework evaluates trim, maneuver performance, and lateral obstacle avoidance. Results show morphing wings significantly expand the flight envelope by decoupling lift and pitch requirements. In dynamic maneuvers, morphing yields distinct trade-offs: a pull-up maneuver increased altitude gain by 28.95\% at a higher control cost, while a banked turn improved lateral displacement by 8.62\% while reducing control cost by 13.40\%. Notably, in obstacle avoidance, morphing reduced total control cost by 65.65\%. This efficiency stems from exploiting aero-mechanical coupling via trajectory optimization to identify coordinated control strategies that offload aerodynamic loads. These findings underscore wing morphing’s potential for achieving extreme maneuvers with superior energy efficiency.
\end{abstract}

\section*{Nomenclature}

{\renewcommand\arraystretch{1.0}
\noindent\begin{longtable*}{@{}l @{\quad=\quad} l@{}}
  $\mathbf{A}$ & Aerodynamic influence matrix \\
  $\mathbf{a}$ & Vector of aerodynamic states \\
  $\dot{\mathbf{a}}$ & Time derivative of aerodynamic states \\
  $\mathbf{B}$ & Jacobian of joint constraints \\
  $b$ & Wing span, m \\
  $\mathbf{C}$ & Wake convection matrix \\
  $c$ & Chord length, m \\
  $\mathbf{F}$ & State-space system dynamics \\
  $\boldsymbol{F}$ & Force vector, N \\
  $\mathbf{f}$ & Structural force vector \\
  $\mathbf{f}_{\text{aero}}$ & External aerodynamic force vector \\
  $\mathbf{G}$ & Effective flow velocity mapping function \\
  $\mathbf{g}$ & Constraint functions \\
  $\mathbf{H}$ & Aerodynamic force mapping function \\
  $h$ & Altitude, m \\
  $h_{obs}$ & Obstacle height, m \\
  $J$ & Total cost function \\
  $L$ & Transient cost function \\
  $\mathbf{M}$ & Mass matrix \\
  $\boldsymbol{M}$ & Moment vector, N$\cdot$m \\
  $\mathcal{O}$ & Obstacle region space \\
  $P$ & Actuation power, W \\
  $q$ & Pitch rate, rad/s \\
  $\mathbf{q}$ & Generalized coordinates \\
  $\mathbf{q}_e$ & Elastic coordinates \\
  $\mathbf{q}_r$ & Rigid coordinates \\
  $R_{eff}$ & Effective radius, m \\
  $R_{obs}$ & Obstacle radius, m \\
  $T$ & Final time of trajectory, s \\
  $t$ & Instantaneous time, s \\
  $t_s, t_f$ & Start and end times of evaluation window, s \\
  $\mathbf{u}$ & Total control vector \\
  $u$ & Forward velocity, m/s \\
  $\mathbf{u}_c$ & Conventional control vector \\
  $\mathbf{u}_m$ & Morphing control vector \\
  $\mathcal{U}$ & Admissible control set \\
  $W$ & Work, J \\
  $w$ & Vertical velocity, m/s \\
  $w_{\chi}$ & Weighting factors in trim cost function \\
  $\mathbf{x}$ & State vector \\
  $\mathcal{X}$ & Admissible state set \\
  $x, y, z$ & Inertial position coordinates, m \\
  $\alpha$ & Angle of attack, rad \\
  $\beta$ & Dihedral or winglet deflection angle, rad \\
  $\delta$ & Control deflection, rad \\
  $\epsilon$ & Regularization constant \\
  $\boldsymbol{\lambda}$ & Lagrange multipliers \\
  $\Phi$ & Terminal cost function \\
  $\phi, \theta, \psi$ & Roll, pitch, and yaw angles, rad \\
  $\rho$ & Air density, kg/m$^3$ \\
  $\sigma^2$ & Oscillation metric (variance) \\
  $\hat{\chi}$ & Normalized state variable \\
  \multicolumn{2}{@{}l}{Subscripts}\\
  $A$ & Aileron \\
  $E$ & Elevator \\
  $eff$ & Effective values \\
  $goal$ & Target/Goal state \\
  $M$ & Morphing \\
  $o$ & Initial or trim condition \\
  $R$ & Rudder \\
  $T$ & Thrust \\
  \multicolumn{2}{@{}l}{Superscripts}\\
  $\top$ & Matrix transpose \\
  $\cdot$ & Time derivative \\
\end{longtable*}}

\section{Introduction} \label{introduction}

Morphing aerial vehicles, which can change their physical configurations in flight, offer significant advantages over non-morphing counterparts, including drastically increased maneuverability \cite{Barbarino2011,Harvey2022,li2018review, Shepard2016}, energy efficiency and durability, and multifunctionality and task versatility \cite{Ajanic2022, Jeger2024, Vourtsis2023}.

Morphing wing technologies are broadly classified into airfoil-level (2D) and wing-level (3D) concepts \cite{li2018review}. 
Airfoil-level morphing involves localized modifications to the wing's cross-section, such as varying 
the camber at the leading and trailing edges to control lift distribution \cite{monner2009design, monner2001realization}, or altering the airfoil's 
thickness to reduce drag by delaying the laminar-to-turbulent flow transition \cite{courchesne2010new}. In contrast, wing-level 
morphing involves large-scale changes to the entire wing planform to accommodate different phases of 
flight, such as takeoff, cruise, and landing. Example strategies include span-wise morphing to balance 
fuel efficiency with maneuverability \cite{vale2011aero},
variable sweep to enable effective flight across different 
speed regimes \cite{laursen2008robot}, twist morphing for load alleviation and roll control \cite{jenett2017digital}, and folding wing concepts that 
alter wing area to enhance stability and flight characteristics \cite{mills2017flight}.

Particularly, one of the major potential advantages of morphing aircraft is to achieve the high agility and maneuverability that are similar to birds. Consequently, there has been significant research interest in the development of morphing configurations that maximize this benefit \cite{ajaj2021recent}. A key prerequisite of such maximization is the quantification of the enhancement in aircraft maneuverability due to morphing. 

Numerous studies have used static or quasi-static analysis to demonstrate the potential for morphing to enhance aircraft maneuverability. In Mills et al. \cite{mills2017flight}, it was experimentally demonstrated that folderons can actively modify lateral and directional stability: symmetric upward deflection increases dihedral effect, enhancing lateral stability; symmetric downward deflection induces anhedral, reducing stability; and asymmetric deflections generate drag differentials for yaw control, enabling more agile maneuvers by temporarily reducing stability. Bradley et al. \cite{bradley2024structural} performed a trade study comparing the maneuverability benefit of various morphing modes (camber, dihedral, span, sweep, and taper) using panel methods to develop effectiveness and sensitivity metrics, analogous to stability derivatives. These metrics were derived from quasi-steady perturbations around trim conditions to evaluate the instantaneous aerodynamic authority of each morphing mode. 

However, to properly assess the maneuverability benefits of morphing, it is arguably insufficient to employ only a quasi-steady aerodynamic analysis focusing on changes in lift, drag, and moment coefficients. Instead, a comprehensive evaluation requires a model that captures the full dynamic effects of the morphing process itself. Rapid configuration changes can induce significant unsteady aerodynamic phenomena, such as dynamic stall and complex wake interactions, which lead to transient variations in aerodynamic loads. Furthermore, the structural response to these loads and the morphing actuation introduces critical aeroelastic and aeroservoelastic effects, including the potential for control reversal. These dynamic interactions can fundamentally alter the vehicle's stability and control authority \cite{ajaj2021recent}. Therefore, a multi-physics model capable of capturing these coupled transient aeroservoelastic and morphing effects is essential to fully characterize the impact of morphing on aircraft maneuverability.

Furthermore, a quantitative comparison of maneuverability between morphing and conventional non-morphing vehicles requires not only the multi-physics dynamics model, but also its coupling to trajectory optimization. This is because morphing aircraft, which inherently have highly nonlinear dynamics, necessitate that the full vehicle dynamics be considered to generate dynamically feasible paths \cite{smith2025parametric}. By generating optimal flight paths for both configurations under identical performance criteria and constraints, the trajectory optimization allows for the direct calculation of key performance metrics in the optimal sense. The benefits of morphing can then be quantified through metrics such as maximum altitude gain, minimum turn radius, or enhanced obstacle avoidance capabilities, that are potentially more comprehensive than those based on linearized dynamics at specific trim states (e.g., \cite{mills2017flight, bradley2024structural}).

Towards the desired aeroservoelastically-coupled trajectory optimization, three key components must be integrated: (1) high fidelity aeroservoelastic analysis methods, (2) full 6-degrees-of-freedom (DOF) flight dynamics, and (3) a trajectory optimization framework. There have been research efforts in integrating two of components, but much fewer efforts that integrate all three.

One example focuses on aeroelasticity and trajectory analysis, with lower-fidelity models for the sake of computational cost \cite{cervantes2024development, phillips2022three}.
In Ref.~\cite{cervantes2024development}, the SPARRO (Software for Parallelized Analyses, Reconfiguration, and Rapid Optimization) framework is developed and used a low-fidelity preprocessor to screen the design space by optimizing static performance for discrete mission segments (such as cruise or loiter). Higher-fidelity analyses are then applied only to these pre-selected configurations. Complementary to the integrated design is the development and validation of computationally efficient aeroelastic analysis methods such as the Uncoupled Static Aeroelastic Analysis (U-SAA) \cite{phillips2022three}, a decoupled aeroelastic analysis methods which uses surrogate models to accurately predict static aeroelastic response. However, these methods do not consider the 6-DOF flight dynamics required to properly assess the benefits of morphing in dynamic maneuvering.

In parallel, a second thrust has concentrated on the development of time-domain simulation frameworks that integrate aeroelasticity and flight dynamics. For example, a mid-fidelity non-linear, time-domain, multibody simulation framework was developed by Cheng et al. \cite{cheng2025nonlinear} for the analysis of coupled aeroelasticity and flight dynamics of flexible aircraft with passive, flared hinged wingtips.  Preston et al. \cite{preston2025multibody} extended this framework for the simulation of dynamically actuated wingtips and showed that variable sweep wingtips can offer a dual benefit of gust load alleviation and active flight control for roll and pitch maneuvers. In the framework, 6-DOF flight dynamics was coupled with a unsteady vortex lattice method (UVLM) for aerodynamics and a non-linear geometrically exact beam theory (GEBT) for structure. For structural dynamics, the connections between bodies are modeled through a Lagrange multiplier formulation. However, the models have not yet been integrated into a trajectory optimization framework to assess the effect of morphing.

The third thrust attempts to integrate aeroelastic analysis methods and a dynamic simulation platform for trajectory optimization, in order to eventually perform simultaneous design and trajectory optimization to fully leverage the benefits of morphing for mission-specific requirements. For example, Jasa et al. \cite{jasa2018design} simultaneously optimized the design and trajectory of a morphing Common Research Model (CRM) for fuel burn, demonstrating that continuous wing twist changes could yield a 0.2\% to 0.7\% reduction in fuel burn compared to a non-morphing design, by coupling a rigid-body flight dynamics model with a quasi-steady Vortex Lattice Method (VLM) for aerodynamics and locally linear 6-DOF spatial beam elements for the structure. More recently, Bakhshi et al. \cite{bakhshi2025coupled} examined the benefits of wing morphing for a field-deployed UAV using a collocation-based trajectory optimization directly coupled to an aerostructural simulation. The results demonstrated that modular morphing can significantly enhance mission-level performance, achieving an 18.29\% reduction in optimal trajectory time for aerodynamics-only cases and an 11.72\% reduction when structural failure constraints were included. However, to remain computationally tractable, their study relied on low-fidelity quasi-steady aerodynamics that neglected unsteady effects and simplified the problem by considering only quasi-constant-altitude flight, omitting 6-DOF flight dynamics. Furthermore, in Ref. \cite{bakhshi2025coupled}, the control cost was not accounted for in the optimization, and the trade-off between maneuverability and the required actuation effort remains unclear.

The above discussion highlights an unaddressed gap: a trajectory optimization framework that integrates a sufficiently high-fidelity, unsteady aeroservoelastic model with 6-DOF flight dynamics, while explicitly accounting for accurate physics-based control costs. Such a framework would enable a rigorous quantification of the maneuverability benefits of morphing aircraft, capturing not only the complex interactions between morphing actuation, unsteady aerodynamics, structural dynamics, and flight control, but also the critical trade-off between performance gains and the required actuation effort. Our efforts aim to fill this gap by progressively building and integrating these essential components.

Our prior work \cite{smith2025parametric} performed a preliminary study on the effect of morphing on aerial vehicles in maneuverability and obstacle avoidance scenarios, based on a low-fidelity linear parameter-varying model. In the study, it was discovered that the ability to morph dihedral angles greatly increased the maneuverability in longitudinal dynamics (e.g., pull-up) and horizontal dynamics (e.g., turn) via a reachability analysis. Furthermore, it was found that trajectory \textit{optimization} is necessary to achieve the desired maneuverability in obstacle avoidance scenarios, instead of the more commonly used trajectory \textit{planning}. The key difference between trajectory optimization and planning is that the latter considers only the vehicle kinematics and not the complete vehicle dynamics. For simpler autonomous systems, such as quad-rotor drones, that are differentially flat~\cite{van1998real}, one would only need path planning to generate a kinematically feasible trajectory and then infer the required control. Yet, due to the high nonlinearity of morphing vehicle dynamics, a planned trajectory might require excessively high control effort, or even be dynamically infeasible, for the vehicle to track.  Hence, an optimization process that considers the dynamical constraints is needed to produce a dynamically feasible trajectory with a sufficiently low control effort. Nevertheless, in Ref.~\cite{smith2025parametric}, the model was of lower fidelity and did not consider the unsteady aerodynamic effects due to morphing and the aeroelastic effects on the wing. 

Our subsequent research \cite{pudasaini2025aeroservoelastic} continued to develop a higher-fidelity aeroservoelastic model for an aircraft with folding winglets, combining GEBT and UVLM. The model was integrated into a trajectory optimization framework to analyze a pull-up maneuver. The results showed that morphing winglets can substantially improve performance in a pull-up maneuver, achieving a 9.9 \% increase in altitude gain. However, there were still several missing aspects. The morphing joint was modeled as a short wing segment of reduced stiffness instead of a more realistic articulated joint. Furthermore, the optimization neglected control cost, failing to account for the crucial trade-off between maneuverability and the required actuation effort. Finally, the framework was only applied to a pull-up maneuver, leaving more complex scenarios such as turns and obstacle avoidance unexplored.

This paper continues the above streak of effort, and aims to develop and apply an integrated trajectory optimization framework, incorporating a high-fidelity aeroservoelastic model and a detailed control cost model, to assess the dynamical maneuverability of morphing winglet aircraft. The specific objectives of this paper are as follows.
    \begin{enumerate}
        \item Present an aeroservoelastic model for an aircraft with morphing winglets, utilizing a multi-body dynamic formulation to represent articulated winglet joints.
        \item Integrate the aeroservoelastic model and a physics-based control cost model in a trajectory optimization framework.
        \item Apply the framework to quantify the trajectory-level trade-off between performance gains and control costs due to morphing across various scenarios, including typical maneuvers, reachability analysis, and obstacle avoidance.
    \end{enumerate}

\section{Methodology}

\subsection{Model Configuration}

\begin{figure}[hbt!]
    \centering
    \includegraphics[width=\linewidth]{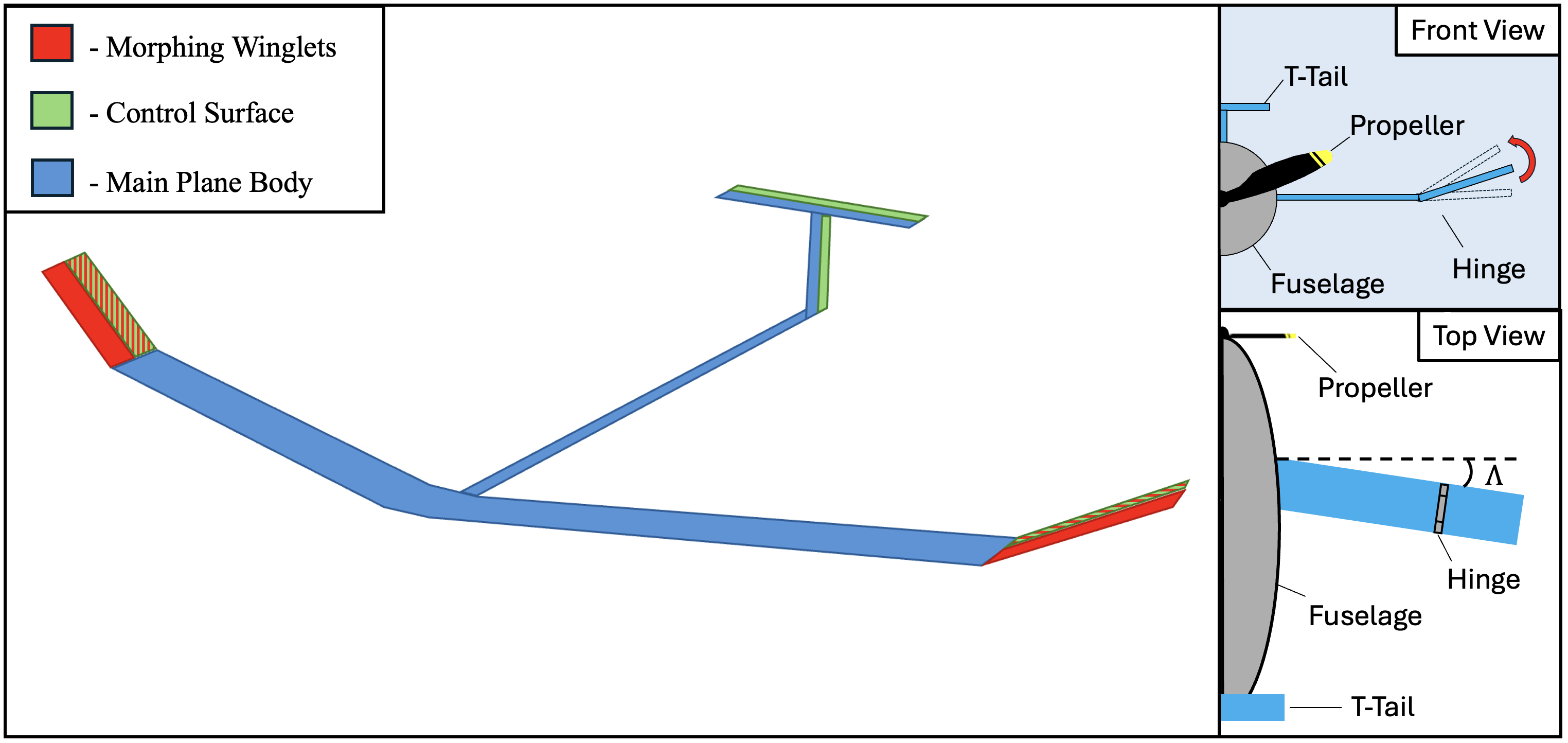}
    \caption{Morphing aircraft configuration with folding winglets.}
    \label{config}
\end{figure}

This work considers a vehicle configuration derived from the High Altitude Long Endurance (HALE) 
model~\cite{murua2010modeling, wang2006time}. The model features flexible, high-aspect-ratio wings and a rigid 
T-tail, as illustrated in Fig.~\ref{config}. The outer segments of the wing, referred to as winglets, 
are assumed to be morphing, and can vary their dihedral angles by rotating around the hinge.

\subsection{Aircraft Dynamics}

The configuration used in the study involves 
the nonlinear interaction among the unsteady aerodynamics, the geometrically nonlinear elastic effects, 
and the morphing behaviors. To capture such nonlinear dynamics, the well-validated, open-source package 
SHARPy \cite{del2019sharpy,preston2025multibody} is used.  It contains the following components:

\begin{enumerate}
    \item Unsteady Vortex Lattice Method (UVLM) to model unsteady aerodynamics with wake effects;
    \item Geometrically Exact Beam Theory (GEBT) for geometrically nonlinear wing deflections and rotations; 
    \item Six degree-of-freedom (6 DoF) flight dynamics with capability to actuate control surfaces and point forces;
    \item Multi-body dynamics using Lagrange-based constraints for articulated joints.
\end{enumerate}

The particular SHARPy model provided in Ref.~\cite{preston2025multibody} is used in this study.  The overall aeroservoelastic system is:
\begin{subequations}
    \label{eqn_coupled_system}
    \begin{align}
        \mathbf{M}(\mathbf{q}, \mathbf{u}_m)\ddot{\mathbf{q}} + \mathbf{f}(\mathbf{q}, \dot{\mathbf{q}}, \mathbf{u}_m) 
        + \mathbf{B}^{\top}(\mathbf{q}, \dot{\mathbf{q}}, \mathbf{u}_m)\boldsymbol{\lambda} 
        &= \mathbf{f}_{\text{aero}}, \label{eqn_struct} \\[2mm]
        \mathbf{g}(\mathbf{q}, \dot{\mathbf{q}}, \mathbf{u}_m) &= \mathbf{0}, \label{eqn_constraints} \\[1mm]
        \mathbf{A}(\mathbf{q}, \mathbf{u}_m) \dot{\mathbf{a}} + \mathbf{C}(\mathbf{q}, \mathbf{u}_m) \mathbf{a} 
        &= \mathbf{G}(\mathbf{q}, \dot{\mathbf{q}}, \ddot{\mathbf{q}}, \mathbf{u}_c, \mathbf{u}_m), \label{eqn_aero} \\[1mm]
        \mathbf{f}_{\text{aero}} 
        &= \mathbf{H}(\mathbf{q}, \dot{\mathbf{q}}, \ddot{\mathbf{q}}, \mathbf{a}, \dot{\mathbf{a}}, \mathbf{u}_c, \mathbf{u}_m). \label{eqn_aero_force}
    \end{align}
\end{subequations}

Equations \eqref{eqn_struct} and \eqref{eqn_constraints} are the structural dynamics equations, where \( \mathbf{q} \) is the vector of generalized coordinates describing the global attitude, rigid-body velocities, and flexible deformations of the multibody structure; \(\mathbf{g}\) is the joint constraint.  Here, \(\mathbf{M}\) is the mass matrix, \(\mathbf{f}\) includes the gyroscopic and stiffness forces, and \(\mathbf{f}_{\text{aero}}\) is the external aerodynamic force. The matrix \(\mathbf{B} = \partial \mathbf{g}/\partial \dot{\mathbf{q}}\) is the Jacobian of the joint constraints, and \(\boldsymbol{\lambda}\) are the Lagrange multipliers enforcing the constraints \( \mathbf{g} = \mathbf{0}.\)

Equations \eqref{eqn_aero} and \eqref{eqn_aero_force} are the aerodynamic equations of motion, where $\mathbf{a}$ is the vector of aerodynamic states, including bound vortex strengths and wake vortex strengths and positions. Here, \(\mathbf{A}\) is the aerodynamic influence matrix, \(\mathbf{C}\) is the wake convection matrix, \(\mathbf{G}\) is the effective flow velocity vector, and \(\mathbf{H}\) is the mapping from aerodynamic states to aerodynamic forces.

Control inputs are separated into morphing inputs, $\mathbf{u}_m$, which modify the structural and aerodynamic properties and impact all terms, and conventional control surface deflections, $\mathbf{u}_c$ (elevator, ailerons, rudder), which only modify aerodynamics through $\mathbf{G}$.

Defining the combined state $\mathbf{x} = [\mathbf{q}^{\top}, \dot{\mathbf{q}}^{\top}, \mathbf{a}^{\top}]^{\top}$ and control $\mathbf{u} = [\mathbf{u}_c^{\top}, \mathbf{u}_m^{\top}]^{\top}$, the state-space form of Eq.~\eqref{eqn_coupled_system} is
\begin{align}
    \mathbf{F} (\mathbf{x}, \dot{\mathbf{x}}, \mathbf{u}) &=
    \begin{bmatrix}
        \mathbf{M}\ddot{\mathbf{q}} +  \mathbf{f} + \mathbf{B}^{\top} \boldsymbol{\lambda} - \mathbf{f}_{\text{aero}} \\[1mm]
        \mathbf{A}\dot{\mathbf{a}} + \mathbf{C} \mathbf{a} - \mathbf{G} \\[1mm]
        \mathbf{g}
    \end{bmatrix} = \mathbf{0}
\end{align}

Both the aerodynamic term $\mathbf{G}$ and the external aerodynamic force $\mathbf{f}_{\text{aero}}$ generally depend on the structural acceleration $\ddot{\mathbf{q}}$. Therefore, the system is a Differential-Algebraic Equations (DAEs), that needs to be solved by an implicit time integration method \cite{del2019sharpy}.

\subsection{Trim}

In the preliminary phase of aircraft trajectory optimization, one needs to determine a reference equilibrium point of the aircraft (trim conditions), which constitute the initial conditions for the optimization process. This step is particularly critical for aeroservoelastic dynamics, as untrimmed aircraft states can result in significant elastic response in the initial transients and thus spurious oscillations in the flight dynamics.

For the ease of discussion below, the generalized structural coordinates \(\mathbf{q}\) is further partitioned as
\( \mathbf{q} = [\mathbf{q}_r^{\top}, \mathbf{q}_e^{\top}]^{\top}.
\) Here, \(\mathbf{q}_r = [x, y, z, \phi, \theta, \psi]^{\top}\) are the rigid-body coordinates, consisting of  the inertial positions \( (x, y, z) \) and the Euler angles \( (\phi, \theta, \psi) \), and \(\mathbf{q}_e\) are the elastic coordinates, consisting of the translational and rotational degrees of freedom of the structural nodes in the body-attached frame of reference.

Trimming an aircraft is the process of determining a specific flight state \( \mathbf{x}_o = [\mathbf{q}_{r_o}^{\top}, \mathbf{q}_{e_o}^{\top}, \dot{\mathbf{q}}_{r_o}^{\top}, \dot{\mathbf{q}}_{e_o}^{\top}, \mathbf{a}_o^{\top}]^{\top} \) and a constant input \( \mathbf{u}_o = [\mathbf{u}_{c_o}^{\top}, \mathbf{u}_{m_o}^{\top}]^{\top} \) such that all accelerations and internal rates of change are zero:
\(
    \dot{\mathbf{x}}_o = [\dot{\mathbf{q}}_{r_o}^{\top}, \mathbf{0}^{\top}, \mathbf{0}^{\top}, \mathbf{0}^{\top}, \mathbf{0}^{\top}]^{\top}
\). The trim condition is then found by solving the nonlinear algebraic equations:
\begin{equation}\label{eqn:sta_trim}
    \mathbf{F}(\mathbf{x}_o, \dot{\mathbf{x}}_o, \mathbf{u}_o) = \mathbf{0}
\end{equation}
where the unkowns are the equilibrium state vector \( \mathbf{x}_o \) and the control input vector \( \mathbf{u}_o \).

For this work, the trim in the longitudinal direction is considered, and the variables simplify to: 
\[
    \mathbf{q}_r = [x, z, \theta]^{\top}, \quad
    \dot{\mathbf{q}}_r = [u, w, q]^{\top}, \quad
    \mathbf{u} = [\delta_E, \delta_M, \delta_T]^{\top}
\]
where the relevant rigid body coordinates are the horizontal position \( x \), the vertical position \( z \), and the pitch angle \( \theta \), and the control vector reduces to the inputs affecting longitudinal motion only, including elevator deflection \( \delta_E \), winglet dihedral \( \delta_M \) and differential thrust \( \delta_T \). The trim condition corresponds to steady, level flight, characterized by constant horizontal velocity \( u_0 \) and zero vertical speed \( w_0 \) and pitch rate \( q_0 \):
\(
    \dot{\mathbf{q}}_{r_o} = [u_o, 0, 0]^{\top}
\).

\subsubsection{Conventional Trimming Procedure}
Aircraft trim is typically obtained by iteratively solving the steady aeroelastic equilibrium defined by Eq.~\eqref{eqn:sta_trim} using gradient-based methods. 

However, for flexible aircraft with complex multibody features such as hinged winglets, the hinge mechanism introduces strong nonlinearities and sensitivity that can lead to poor convergence or nonconvergence when using a purely static algebraic approach \cite{cheng2025nonlinear}. Moreover, even when a static equilibrium is found, it may be dynamically unstable, as the algebraic approach neglects the unsteady coupling between the flexible structure and the developing aerodynamic wake.

\subsubsection{Hybrid Static-Dynamic Procedure}
To address the convergence and stability limitations of the conventional static trimming approach, a hybrid static-dynamic methodology was adopted. First, a static trim analysis provides a rapid initial estimate of the equilibrium state, which is subsequently refined through a dynamic trim optimization.

\paragraph{Static Trim} 
Initial estimates for the angle of attack $\alpha$ and controls $\mathbf{u} = [\delta_E, 0, \delta_T]^\top$ are obtained for a target velocity $u_o$ and a fixed winglet dihedral $\delta_M = 0$ using the \textit{StaticTrim} routine in SHARPy. This routine iteratively solves the nonlinear algebraic system in Eq.~\eqref{eqn:sta_trim} by employing a simplified first-order model, which updates the control inputs based on the physical argument that primary aerodynamic force and moment residuals are predominantly sensitive to specific primary controls \cite{cheng2025nonlinear}. The iterative process attempts to drive the net external loads acting on the deformed structure to zero, although numerical convergence is not always guaranteed.

The aircraft may not be trimmable at the target flight speed under the constraint of a fixed winglet dihedral ($\delta_M = 0$). Furthermore, even if a static equilibrium is found, it may prove dynamically unstable. To address these limitations, a dynamic trim optimization is performed, where the winglet dihedral ($\delta_M$) is incorporated as an additional control degree of freedom to ensure a stable and consistent equilibrium.

\paragraph{Dynamic Trim} 
To achieve a stable equilibrium, an optimal control problem (OCP) is formulated over a short horizon $[0, t_f]$, using the static trim result as an initial guess. The optimization is defined as:
\begin{subequations}
\label{eqn:trim_analysis}
\begin{align}
    & \min_{\alpha, \mathbf{u}} J_\text{trim}\big( \mathbf{q}_r(t; \alpha, \mathbf{u}) \big) \\
    \text{s.t.: } \quad & \mathbf{F}(\mathbf{x}(t; \alpha), \dot{\mathbf{x}}(t; \alpha), \mathbf{u}) = \mathbf{0} \\
    & \mathbf{u}\in \mathcal{U}, \quad \mathbf{x}(t)\in \mathcal{X}
\end{align}
\end{subequations}
where the optimization variables are the control inputs \(\mathbf{u} = [\delta_E, \delta_M, \delta_T]^{\top} \) and the angle of attack \(\alpha\) which affects the reduced rigid body coordinates \(\mathbf{q}_r(t; \alpha, \mathbf{u})\). The \( J_{\text{trim}} \) (detailed in Appendix A) penalizes deviations from the target steady-state trajectory. The trim variables are optimized such that the vehicle is in a steady level flight after $t_s$.

\subsubsection{Practical Implementation} 
In SHARPy simulations, the aircraft always starts from an undeformed structural state. Hence, a special starting procedure is needed to ensure the aircraft to be dynamically trimmed in the time-domain simulation. The simulation is run from the undeformed initial configuration given the optimized trim inputs for a duration of $t_s = 0.5\,\mathrm{s}$. This procedure allows the structural deformation and aerodynamic wake to fully develop and converge to a consistent, fully coupled aeroservoelastic equilibrium. Subsequently, the variations of inputs, e.g., from trajectory optimization, are supplied after $t = t_s$. In other words, to simulate a time-domain response of $T$ seconds, the actual SHARPy simulation will always take $(T + t_s)$ seconds.

\subsection{Trajectory Optimization}

Trajectory optimization is the process of determining an optimal trajectory, i.e.,the sequence of states and control inputs, that is needed to achieve a desired objective while respecting dynamical constraints and operational limits.

Formally, the problem is written as:
\begin{subequations}\label{eqn_to_implicit}
\begin{align}
    \min_{\vu(t)}&\ J = \Phi(\vx(T)) + \int_0^{T} L(\vx(t),\vu(t)) dt \label{eqn_to_obj} \\
    \mathrm{s.t.}&\ \mathbf{F}(\vx(t), \dot{\vx}(t), \vu(t)) = \mathbf{0},\quad t \in [0, T] \label{eqn_dyn_implicit}\\
    &\ \vx(0) = \vx_o, \quad \vu(0) = \vu_o \label{eqn_ic_implicit} \\
    &\ \vx(T) = \vx_f \label{eqn_fc_implicit}\\
    &\ \vu(t)\in \mathcal{U},\ \vx(t)\in \mathcal{X} \label{eqn_cns_implicit}
\end{align}
\end{subequations}
where \(J\) is the total cost, consisting of a terminal cost \(\Phi(\vx(T))\) and a running cost \(L(\vx, \vu)\) integrated over the horizon \(t \in [0, T]\). Equation~\eqref{eqn_dyn_implicit} is the system dynamics, where \(\vx(t)\) and \(\vu(t)\) represent the state and control vectors, respectively. Equation~\eqref{eqn_ic_implicit} specifies the initial trim condition, while Equation~\eqref{eqn_fc_implicit} enforces the terminal condition. Finally, Equation~\eqref{eqn_cns_implicit} constrains the control inputs and states to lie within admissible sets \(\mathcal{U}\) and \(\mathcal{X}\), respectively. The specific forms of cost, terminal condition, and admissible sets are problem-dependent, and will be presented in the results section.

\subsubsection{Actuator Constraints}

To ensure physical feasibility, the admissible control set $\mathcal{U}$ is defined by box constraints on both the control surface deflections and their respective rates of change. These limits, representing typical actuator capabilities, are summarized in Table~\ref{tab:control_bounds}. While the conventional control surfaces (elevator, ailerons, and rudder) follow standard aerospace bounds, the morphing winglets are permitted a significantly larger deflection range of $\pm 60^{\circ}$ to leverage high-authority geometric changes. However, a more conservative rate limit of $\pm 10^{\circ}$/s is applied to the morphing surfaces to account for the increased aerodynamic and structural moments associated with large-scale shape changes. The rate bounds are informed by an estimation of the maximum power output of typical servo motors used in similarly sized model aircraft. 

\begin{table}[hbt!]
    \centering
    \caption{Control surface deflection and rate limits for the trajectory optimization}
    \label{tab:control_bounds}
    \begin{tabular}{lcc} 
        \toprule \toprule 
        \textbf{Control Surface} & \textbf{Deflection Limit (deg)} & \textbf{Rate Limit (deg/s)} \\
        \midrule 
        Elevator ($\delta_E$) & $\pm 15$ & $\pm 20$ \\
        Ailerons ($\delta_A$) & $\pm 15$ & $\pm 20$ \\
        Rudder ($\delta_R$) & $\pm 25$ & $\pm 20$ \\
        Morphing Winglets ($\delta_M$) & $\pm 60$ & $\pm 10$ \\
        \bottomrule \bottomrule 
    \end{tabular}
\end{table}

\subsubsection{Numerical Implementation}

The control input \( \vu(t) \) was discretized using a set of $N$ control points $\mathbf{U} = [\mathbf{u}_1, \mathbf{u}_2, \cdots, \mathbf{u}_N]$, with linear interpolation applied to compute $\mathbf{u}(t)$ for $t \in [0, T]$. Given $\mathbf{U}$ one can simulate the dynamics \eqref{eqn_dyn_implicit} and obtain $\mathbf{x}(t)$, and subsequently evaluate the objective \eqref{eqn_to_obj}. The optimization problem was then solved using the gradient-based 
Sequential Least Squares Quadratic Programming (SLSQP) method, using Python's \textit{SciPy} package. Since analytical gradients are unavailable in \textit{SHARPy}, the gradients of the cost function with respect to the control inputs were approximated using the forward difference scheme. The scheme is implemented in a parallel computing manner, so that the trajectory optimization problem becomes computationally feasible.

\subsection{Control Cost}

Generally, for assessing the maneuverability of an aircraft, the objective is to find a control input sequence that minimizes the total control cost, while satisfying the control constraints (rate limits), final state constraint (reach a goal state), and/or intermediate state constraints (avoid obstacles).

In many trajectory optimization studies, the control cost has been chosen to be quadratic for the sake of simplicity, typically defined as:
\[
    J = \int_{0}^{T} \mathbf{u}(t)^\top \mathbf{R} \mathbf{u}(t) \, dt
\]
where $\mathbf{u}(t)$ is the control input vector and $\mathbf{R}$ is a positive-definite weight matrix. The quadratic cost implies a linear relationship between control inputs and actuation effort, assuming that the cost of deflection is symmetric and independent of the vehicle's state, which is typically valid for conventional control surfaces. 

However, for the morphing control surfaces with large deflections, this relationship is inherently nonlinear and state-dependent. To address this, we adopt a physics-based cost model \cite{smith2026trajectory}, that directly computes actuation effort from aerodynamic loading using the virtual work principle. 

\begin{figure}[hbt!]
    \centering
    \includegraphics[width=\linewidth]{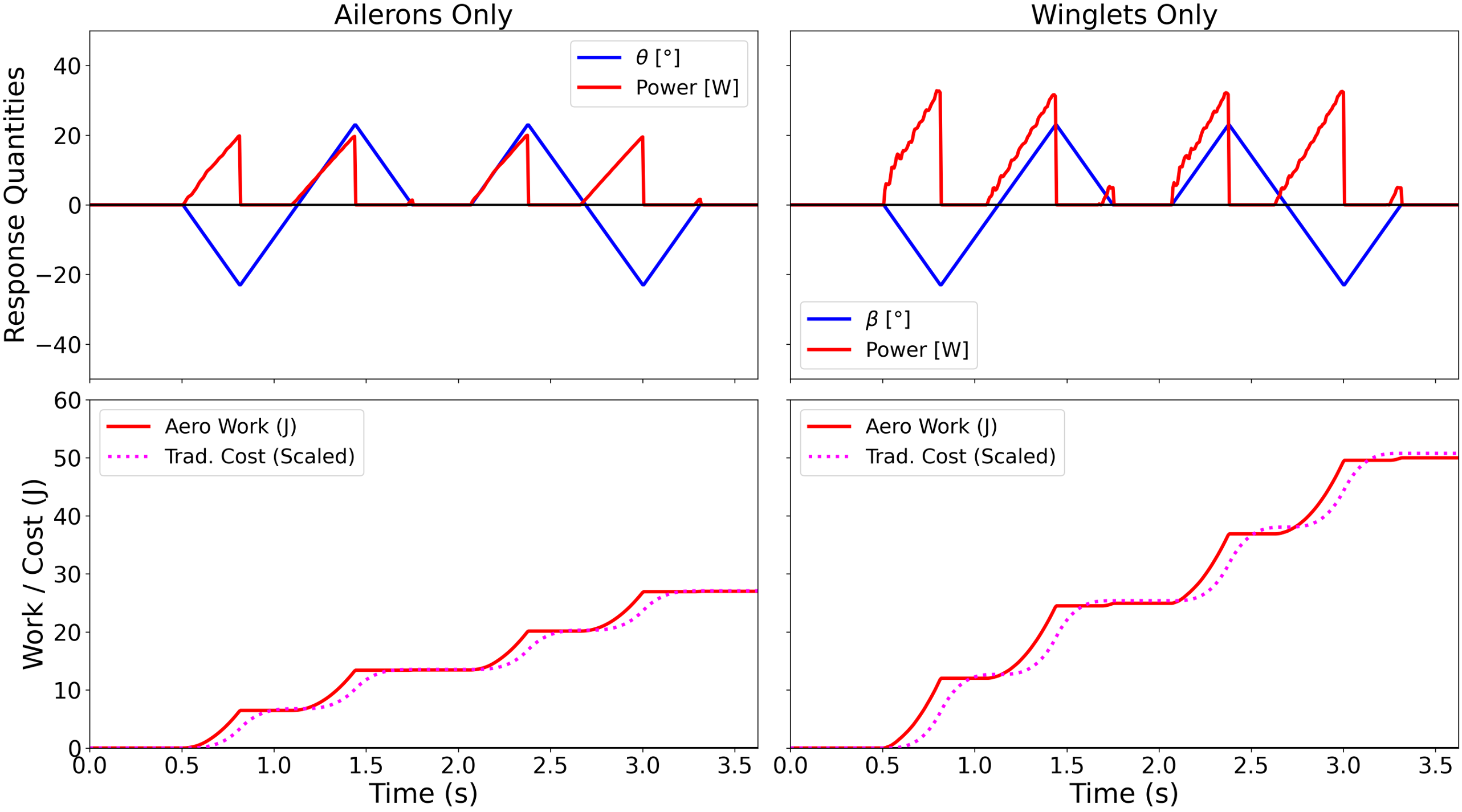}
    \caption{Power analysis of independent control inputs.}
    \label{fig:solo}
\end{figure}

\begin{figure}[hbt!]
    \centering
    \includegraphics[width=\linewidth]{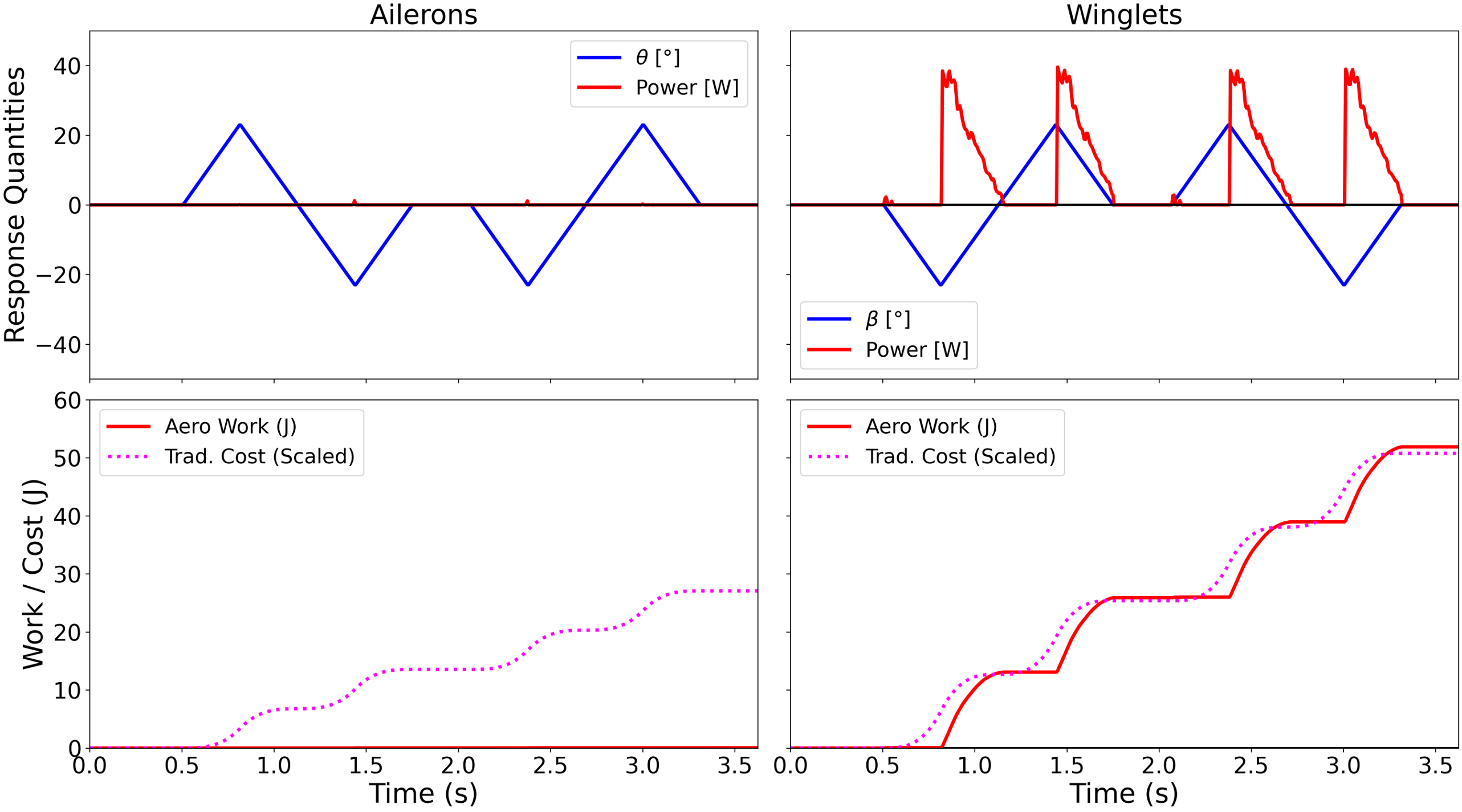}
    \caption{Power analysis of coupled (opposing) control inputs.}
    \label{fig:coupled}
\end{figure}

The necessity of a physics-based cost model is illustrated by the divergence between traditional quadratic cost approximations and the actual aerodynamic work required for maneuver. Figure~\ref{fig:solo} presents the time-history of control surface deflection, instantaneous actuation power, and cumulative work for the aileron and morphing winglet, where each deflection is applied independently. The analysis of independent control surfaces reveals that actuation power is primarily consumed during deflection away from equilibrium to overcome aerodynamic resistance, while the return to neutral is assisted by these accumulated aerodynamic forces. While the quadratic cost captures the general magnitude, it fails to account for this directional bias. Figure~\ref{fig:coupled} presents the time-history of coupled control surface deflections, instantaneous actuation power, and cumulative work for the aileron and morphing winglet when actuated simultaneously in opposing directions. In contrast to the independent case, this demonstrates that coupling the aileron and winglet in opposing directions exploits aerodynamic interactions to cancel resistance. This strategic coordination achieves significant power alleviation for the aileron, yielding near-zero power requirements, which is not captured by the traditional quadratic cost.

\section{Results and Discussion}

The results of this study provide a comprehensive assessment of the effects of morphing winglets on aircraft performance through four analyses of increasing complexity: 
(A) a trim analysis to assess flight envelope expansion; 
(B) a maneuver evaluation comparing pull-up and turn performance; 
(C) a reachability analysis investigating the trade-off between control cost and objective attainment; and 
(D) an obstacle avoidance study demonstrating the advantages of morphing in realistic mission scenarios.

\subsection{Trim Analysis}

The trim analysis defines the aircraft flight envelope by identifying the steady-state equilibrium conditions, thereby establishing the operational limits and initial states required for the maneuver evaluation (B), reachability analysis (C), and obstacle avoidance (D) scenarios.

To assess the capability of morphing aircraft to expand the flight envelope, trim capabilities are evaluated for three configurations. First, the reference configuration is from Ref.~\cite{wang2006time}, and referred to as \textit{elastic fixed-wing}; at trim, it has a wing tip displacement of approximately 14\% of the semi-span, showing significant aeroelastic effects.  Second, as a baseline comparison, a \textit{rigid fixed-wing} configuration is constructed by increasing the wing bending stiffness by 5 times, so that it has a negligible wing tip displacement of approximately 1\% of the semi-span at trim.  Lastly, as the main interest of this study, an \textit{elastic morphing} configuration is based on the elastic fixed-wing configuration but with winglet dihedral angle as an additional control input.

Trim capability is assessed by evaluating whether a configuration can maintain both vertical velocity (${w}$) and pitch rate ($q$) within defined stability tolerances: $|{w}| \leq 0.05$~m/s and $|q| \leq 0.5$~deg/s. As summarized in Table \ref{tab:trim_results}, individual values satisfying stability tolerances are boldfaced, while airspeeds where both criteria are met simultaneously are highlighted with gray shading to indicate a successful steady-state trim condition.

\paragraph{Rigid Fixed-Wing Model} 
As shown in Table~\ref{tab:trim_results}, the rigid configuration trims at only one speed ($30$ m/s). At off-trim speeds, the aircraft faces a fundamental coupling: adjusting the angle of attack ($\alpha$) to maintain lift simultaneously perturbs the pitching moment. While the elevator can compensate for this moment, the resulting change in tail lift prevents simultaneous minimization of ${w}$ and $q$. Consequently, at a low speed of 27.5~m/s, the insufficient lift leads to high thrust commands that violate trim requirements. Conversely, at a high speed of 35~m/s, the increased $\alpha$ and elevator deflection maintain vertical trim but result in a net nose-up moment, failing the $q$ threshold.

\paragraph{Elastic Fixed-Wing Model} 
The elastic fixed-wing model shifts the single trim point to $32.5$ m/s. This increase is attributed to upward tip deflection, which reduces effective span and lift magnitude. To compensate for these aeroelastic losses, a higher airspeed is required to balance the weight. Notably, the elastic wing exhibits reduced $\alpha$-sensitivity to elevator inputs, making $q$ harder to minimize and limiting the aircraft’s ability to trim vertical velocity at higher speeds (e.g., $35$ m/s). 

\paragraph{Elastic Morphing Model} 
The elastic morphing configuration successfully achieves trim across nearly the entire tested range, as illustrated in Table~\ref{tab:trim_results}. By modulating the spanwise lift distribution via winglet deflection, the aircraft effectively decouples lift control from longitudinal attitude. This decoupling allows the winglets to assist in lift generation while the elevator focuses primarily on managing the pitching moment. For instance, at 30~m/s, downward winglet deflection increases the effective span, which enables the elevator to maintain a balanced pitch state. However, trim failure occurs at 25~m/s because the winglets reach their lower actuator bound ($-30^\circ$), resulting in insufficient lift despite a balanced pitching moment. 

Interestingly, the winglets operate more efficiently at higher speeds than at lower speeds. Taking the nominal trim speed of the elastic fixed-wing configuration ($32.5$ m/s) as a reference, the elastic morphing aircraft demonstrates a high-speed margin of $7.5$ m/s compared to a low-speed margin of only $5$ m/s. This asymmetry indicates that morphing is more effective at higher dynamic pressures, where aeroelastic effects are more pronounced.

\begin{table}[hbt!]
    \centering
    \caption{Steady-state trim results across the tested airspeed range where bold values indicate satisfaction of individual stability tolerances ($|w| \leq 0.05$~m/s, $|q| \leq 0.5$~deg/s) and gray-shaded boxes denote successful trim conditions}
    \label{tab:trim_results}
    \renewcommand{\arraystretch}{1.2} 
    \begin{tabular}{llcccccc}
        \toprule \toprule
        \textbf{Configuration} & \textbf{Parameter} & \multicolumn{6}{c}{\textbf{Airspeed (m/s)}} \\
        \cmidrule(lr){3-8}
        & & \textbf{25.0} & \textbf{27.5} & \textbf{30.0} & \textbf{32.5} & \textbf{35.0} & \textbf{37.5} \\
        \midrule
        Rigid Fixed-Wing  & $w$ (m/s)  & -0.4289 & \textbf{-0.0134} & \cellcolor{gray!20}\textbf{0.0003} & 0.2096 & \textbf{0.0259} & 0.6123 \\
                          & $q$ (deg/s) & \textbf{-0.1536} & -1.7398 & \cellcolor{gray!20}\textbf{-0.0230} & \textbf{0.2502} & 5.1937 & \textbf{0.0881} \\
        \addlinespace
        Elastic Fixed-Wing & $w$ (m/s)  & \textbf{0.0027} & \textbf{-0.0013} & \textbf{-0.0072} & \cellcolor{gray!20}\textbf{-0.0250} & 0.1014 & 0.2531 \\
                           & $q$ (deg/s) & -7.2676 & -4.3066 & -1.6493 & \cellcolor{gray!20}\textbf{0.0487} & 0.6148 & 0.6499 \\
        \addlinespace
        Elastic Morphing   & $w$ (m/s)  & -0.1939 & \cellcolor{gray!20}\textbf{-0.0000} & \cellcolor{gray!20}\textbf{-0.0003} & \cellcolor{gray!20}\textbf{0.0027} & \cellcolor{gray!20}\textbf{0.0005} & \cellcolor{gray!20}\textbf{-0.0016} \\
                           & $q$ (deg/s) & \textbf{-0.1908} & \cellcolor{gray!20}\textbf{-0.2167} & \cellcolor{gray!20}\textbf{0.0010} & \cellcolor{gray!20}\textbf{0.0647} & \cellcolor{gray!20}\textbf{0.0019} & \cellcolor{gray!20}\textbf{0.0178} \\
        \bottomrule \bottomrule
    \end{tabular}
\end{table}

\subsection{Study of Typical Maneuvers}

To evaluate the impact of morphing on maneuver performance, two representative maneuvers are considered: (1) a pull-up maneuver; and (2) a turn maneuver. Each maneuver is analyzed for both morphing and non-morphing configurations to quantify the benefits of morphing.

\subsubsection{Pull-Up Maneuver}

\paragraph{Problem Formulation} 
For a pull-up maneuver, the goal is to maximize the final altitude within a fixed time horizon of \( T = 2.0 \text{ sec} \), starting from the trim condition. The maximization is formulated as a trajectory optimization problem as given in Eq.~\eqref{eqn_to_implicit}, where the cost function \( J \) corresponds to the terminal cost \(\Phi(\vx(T))\), defined as the negative final altitude, \( -h(T) \). The transient cost \( L(\vx, \vu) \) is set to zero, as only the final state is of interest in this maneuver. A terminal pitch constraint, $\theta(T) = \theta_{\text{trim}}$, is enforced to ensure a recoverable orientation and prevent unrealistic, high-angle-of-attack behaviors such as near-stall conditions. 

\begin{figure}[hbt!]
    \centering
    \includegraphics[width=\linewidth]{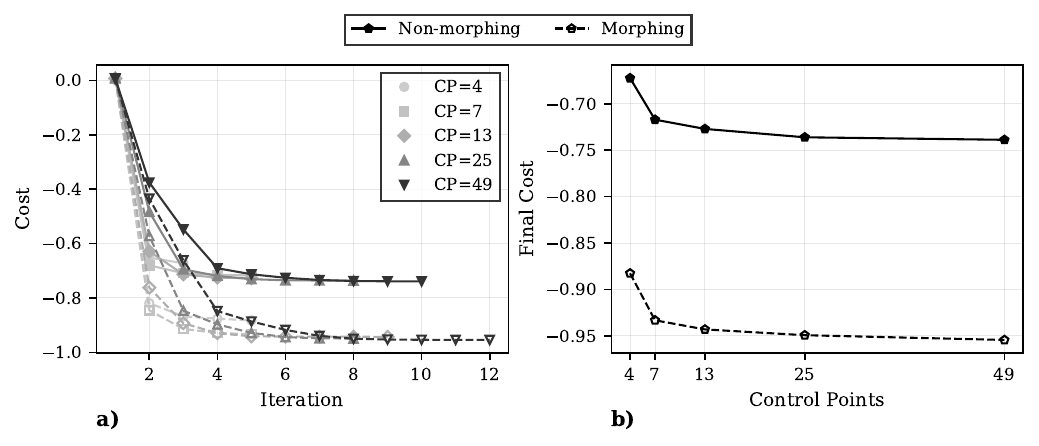}
    \caption{Optimization performance: a) cost convergence history, and b) final cost versus number of control points.}
\label{fig:pullup/cost_analysis_combined}
\end{figure}

To ensure adequate control resolution, a convergence study was performed by increasing the number of 
control points and monitoring the cost function.
Figure~\ref{fig:pullup/cost_analysis_combined}(a) shows the convergence of the cost function over iterations
for different numbers of control points with or without morphing. As seen in Fig.~\ref{fig:pullup/cost_analysis_combined}(b), the final cost value converges at approximately 25 control points. Therefore, this value is used in all subsequent analyses.

To isolate the effects of structural flexibility, the pull-up maneuver was evaluated using both rigid and flexible models, with the rigid non-morphing case serving as the baseline for ideal performance. As shown in Table~\ref{tab:rigid_vs_flexible_performance}, transitioning from a rigid to a flexible non-morphing configuration results in a significant performance drop ($1.2853$~m to $0.7361$~m). This reduction is primarily driven by aeroelastic deformation, where structural compliance results in passive shape changes that reduces control effectivenes. However, the introduction of morphing winglets in the flexible configuration partially recovers this lost capability, yielding a performance gain of $28.95\%$. This suggests that active morphing actuation can be optimized to mitigate the unfavorable effects of aeroelasticity. Furthermore, the performance of the rigid morphing configuration is unrealistically high ($1.6021$~m), as it neglects the structural deformation and aeroelastic damping that naturally limit maneuverability. These results underscore the necessity of accounting for elasticity in the trajectory optimization framework to ensure that the predicted performance gains are physically achievable; consequently, all subsequent analyses are conducted using the flexible aircraft configuration.

\begin{table}[hbt!]
    \centering
    \caption{Altitude gain comparison for rigid and flexible configurations}
    \label{tab:rigid_vs_flexible_performance}
    \begin{tabular}{lccc}
        \toprule \toprule
        \textbf{Model} & \textbf{Non-Morphing (m)} & \textbf{Morphing (m)} & \textbf{Change} \\
        \midrule
        Rigid & 1.2853 & 1.6021 & +24.65\% \\
        \addlinespace
        Flexible & 0.7361 & 0.9492 & +28.95\% \\
        \bottomrule \bottomrule
    \end{tabular}
\end{table}

\paragraph{Control-Response Interpretation}

\begin{figure}[hbt!]
    \centering
    \includegraphics[width=\linewidth]{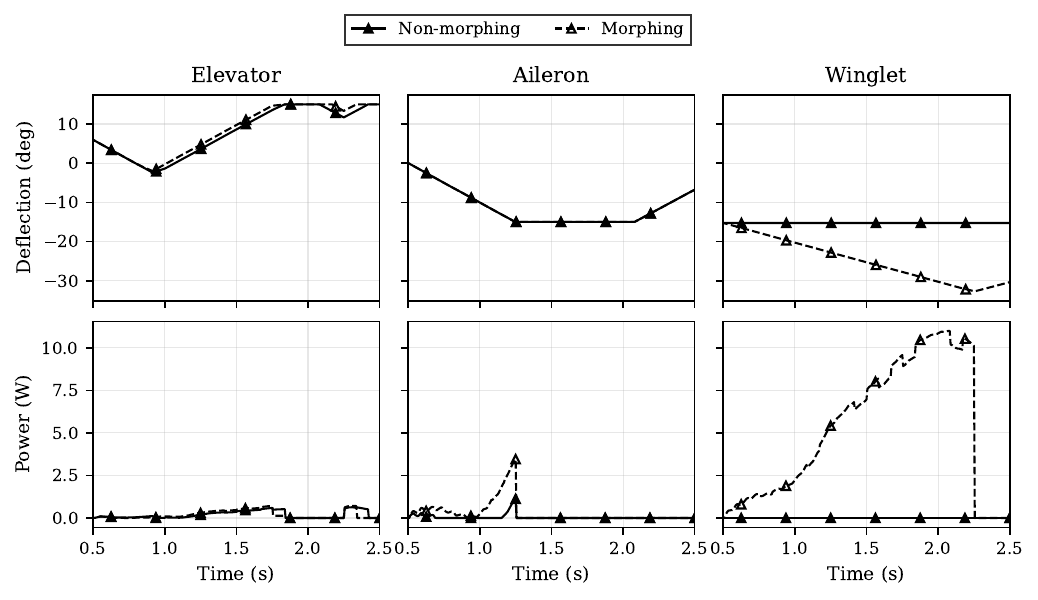}
    \caption{Optimized controls and corresponding power over time for pull-up maneuver.}
    \label{fig:pullup/all_controls_vs_time_single_figure}
\end{figure}

\begin{figure}[hbt!]
    \centering
    \includegraphics[width=\linewidth]{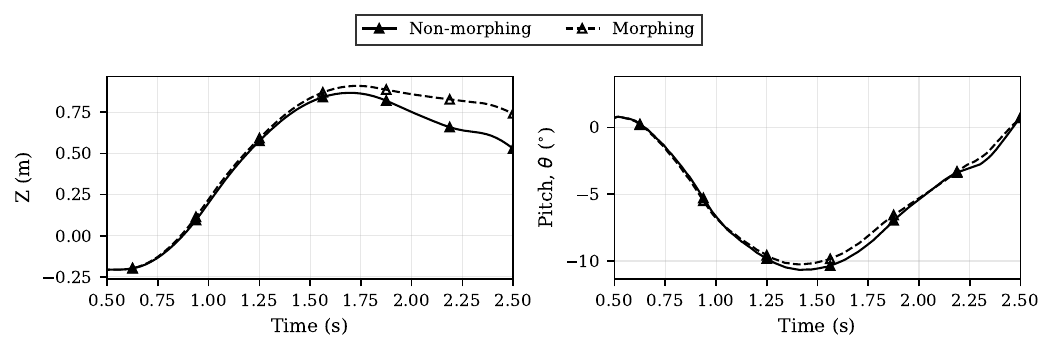}
    \caption{Trajectory over time for pull-up maneuver.}
    \label{fig:pullup/trajectory_vs_time}
\end{figure}

Figure~\ref{fig:pullup/all_controls_vs_time_single_figure} presents the optimized deflections and power consumption for the control surfaces and Fig.~\ref{fig:pullup/trajectory_vs_time} shows the corresponding aircraft response. The optimized control profile follows a consistent and intuitive strategy: it first exploits all available 
control authority to generate maximum lift and rapidly gain altitude, then reverses these actions near the end of the maneuver to restore a near-zero pitch attitude and meet the terminal pitch constraint.

A sharp, downward deflection of the elevator initiates the climb by increasing lift but also induces a strong nose-down pitching moment. To counter this and satisfy the pitch constraint, the elevator deflection is reduced toward the end of the maneuver. Similarly, large downward deflections of ailerons are used early on to enhance lift. In the morphing configuration, the winglets are also deflected downwards, contributing additional lift and resulting in greater altitude gain. Both aileron and dihedral deflections are reduced in the final phase to aid pitch recovery.

\paragraph{Performance and Cost Analysis}

The morphing capability resulted in a notable performance enhancement during the pull-up maneuver, achieving a 28.95\% increase in altitude gain (0.9492~m) relative to the non-morphing configuration (0.7361~m). While the morphing configuration delivered improved performance, it came at a substantially higher control cost. Specifically, the morphing configuration consumed 22.2662~J, due to the winglet actuation, which is approximately 40 times the 0.5513~J required by the non-morphing case. 

\subsubsection{Turn Maneuver}

\paragraph{Problem Formulation}
For a turn maneuver, the goal is to maximize the lateral displacement within a fixed time 
horizon of \( T = 2.0 \ \text{sec} \), starting from the trim condition. In the trajectory optimization problem, the cost
function $ J $ consists of a terminal cost $ \Phi(\mathbf{x}(T)) = -y(T) $, which promotes maximum final lateral position, and a transient cost $ L(\mathbf{x}, \mathbf{u}) = 0 $. A set of terminal state constraint is applied on the altitude, pitch and roll to prevent unrealistic behaviors, such as entering an unrecoverable steep bank:
$$
z(T) = z_{\text{trim}}, \quad \theta(T) = \theta_{\text{trim}}, \quad \phi(T) = 0
$$
where $z_{\text{trim}}$, and $\theta_{\text{trim}}$ are the altitude, and pitch values at trim, respectively.

\paragraph{Control-Response Interpretation}

\begin{figure}[hbt!]
    \centering
    \includegraphics[width=\linewidth]{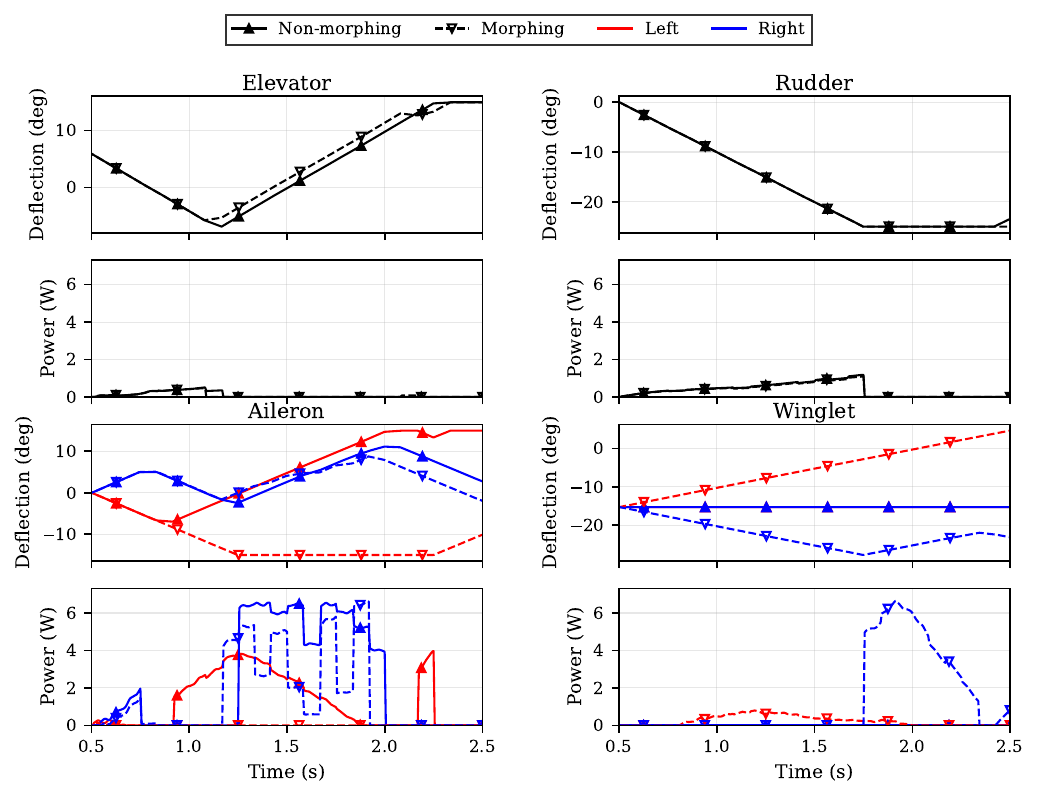}
    \caption{Optimized controls and corresponding power over time for turn maneuver.}
    \label{fig:turn_new/all_controls_vs_time_single_figure}
\end{figure}

\begin{figure}[hbt!]
    \centering
    \includegraphics[width=\linewidth]{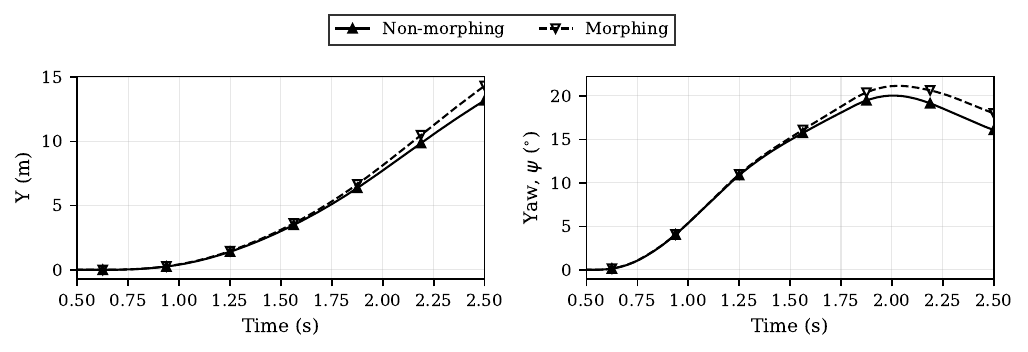}
    \caption{Trajectory over time for turn maneuver.}
    \label{fig:turn_new/trajectory_vs_time}
\end{figure}

Figure~\ref{fig:turn_new/all_controls_vs_time_single_figure} presents the optimized deflections and power consumptions for the control surfaces, while Fig.~\ref{fig:turn_new/trajectory_vs_time} shows the corresponding aircraft response.

The optimized control profile reveals two distinct phases: initiation and recovery. 
During the initiation phase ($t < 1.25 \ \text{s}$), the maneuver begins with a coordinated input. The rudder saturates to a negative angle to induce a strong yawing moment; meanwhile the ailerons deflect anti-symmetrically to bank the aircraft, where the right aileron moves upwards and the left aileron moves downwards. This combination generates the centripetal force required to increase lateral displacement ($y$).
Beyond $t = 1.25 \ \text{s}$, the control strategy shifts to recovery to satisfy the terminal constraints. Notably, in the non-morphing case, the ailerons cease their anti-symmetric behavior and deflect symmetrically (both upwards) alongside the elevator. This suggests they are being utilized to augment lift and pitch authority to level the aircraft. 

In the morphing configuration, the winglets play a critical role in augmenting roll authority. Until $t \approx 1.75 \ \text{s}$, they deflect anti-symmetrically to steepen the bank angle. After $t = 1.75 \ \text{s}$, this anti-symmetric coordination ceases likely to assist in the recovery and stability of the aircraft before the terminal time. Consequently, the morphing aircraft achieves a deeper bank angle ($\phi \approx -14^\circ$) compared to the non-morphing baseline ($\phi \approx -12^\circ$), allowing for a tighter turn radius.

\paragraph{Performance and Cost Analysis}

The morphing aircraft effectively leverages its additional control surfaces to outperform the fixed-wing baseline in both performance and control cost. As summarized in Table~\ref{tab:turn_performance_cost}, the morphing configuration achieves a lateral displacement of $14.3215 \ \text{m}$, representing an $8.62\%$ improvement over the non-morphing case ($13.1844 \ \text{m}$). This gain can be attributed to the increased roll authority provided by the active winglets, which allows the aircraft to bank more aggressively.

\begin{table}[hbt!]
    \centering
    \caption{Performance and control cost comparison for the turn maneuver}
    \label{tab:turn_performance_cost}
    \begin{tabular}{lccc}
        \toprule \toprule
        \textbf{Metric} & \textbf{Non-Morphing} & \textbf{Morphing} & \textbf{Change} \\
        \midrule
        Lateral Displacement (m) & 13.1844 & 14.3215 & +8.62\% \\
        \addlinespace
        Total Control Cost (J) & 7.8778 & 6.9469 & -13.40\% \\
        \bottomrule \bottomrule
    \end{tabular}
\end{table}

\begin{figure}[hbt!]
    \centering
    \includegraphics[width=0.5\linewidth]{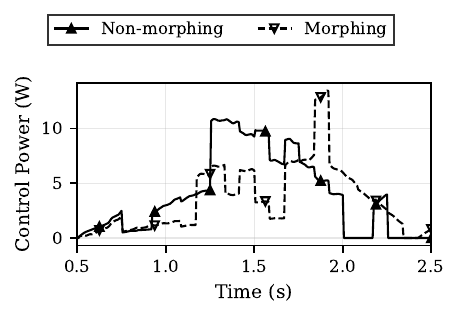}
    \caption{Power consumption over time for turn maneuver.}
    \label{fig:turn_new/control_power_vs_time}
\end{figure}

Despite the morphing configuration exhibiting higher peak power requirements in Fig.~\ref{fig:turn_new/control_power_vs_time} (exceeding $13 \ \text{W}$ compared to the $10 \ \text{W}$ in the baseline), the total control cost for the entire trajectory reduces by $13.40\%$. 
This cost saving is largely attributed to the coupling of the aileron and winglet deflections, which lowers the resistive forces encountered by both surfaces as discussed previously in the control cost section. As seen in Fig.~\ref{fig:turn_new/all_controls_vs_time_single_figure}, the aircraft leverages this ``cancellation effect'' during the initial phase: ailerons and winglets deflect in opposition to minimize net aerodynamic resistance. Subsequently, the ailerons provide static load alleviation, reducing the hinge moments encountered by the winglets. This coordination prioritizes the lower-torque ailerons to mitigate the higher actuation work required by the winglets, reducing the overall control cost.

\subsection{Reachability Analysis}

While lateral turn maneuver demonstrates improved performance alongside reduced control costs, the longitudinal pull-up maneuver achieves performance gains only with a substantial cost penalty. This contrast motivates a further assessment of the benefits of morphing for the longitudinal case. Specifically, a reachability analysis is employed to quantify the expansion of the vehicle's operational envelope.

\paragraph{Problem Formulation}
For the reachability analysis of the pull-up maneuver, the goal is to reach a specified final altitude, $z_{\text{goal}}$, at a trim pitch attitude, $\theta_{\text{trim}}$, within a fixed horizon of $T = 2.0$~s, starting from steady-level flight. This formulation shifts the objective from altitude maximization to cost minimization by treating $z_{\text{goal}}$ as a hard terminal constraint rather than a performance target. This allows for a direct comparison of actuation efficiency at identical setpoints and explicitly identifies the feasibility boundaries where the baseline aircraft fails to satisfy the required terminal state.

The problem is formulated as a trajectory optimization task (Eq.~\eqref{eqn_to_implicit}) minimizing the control work $J = \int L(\mathbf{x}, \mathbf{u}) dt$. The terminal cost $\Phi(\mathbf{x}(T))$ is set to zero, and strict equality constraints are enforced on the final state:
\begin{equation}
z(T) = z_{\text{goal}}, \quad \theta(T) = \theta_{\text{trim}}
\end{equation}
Three distinct target altitudes, $z_{\text{goal}} \in \{0.5, 0.7, 0.9\}$~m, were investigated to quantify the attainable mission space.

\paragraph{Performance and Cost Analysis}

The control cost for these cases is summarized in Table~\ref{tab:reachability_costs}. A critical distinction is observed in the most demanding case ($z_{\text{goal}} = 0.9$~m): the morphing configuration successfully executes the maneuver, whereas the non-morphing baseline fails to reach the target state. For the intermediate cases ($0.5$~m and $0.7$~m), the control cost for the morphing aircraft is equal to that of the non-morphing baseline. This demonstrates that morphing not only extends the flight envelope but also provides the operational flexibility to remain inactive during less demanding tasks, ensuring that the added authority does not result in an efficiency penalty during standard maneuvers.

\begin{table}[hbt!]
    \centering
    \caption{Control cost comparison for pull-up reachability analysis}
    \label{tab:reachability_costs}
    \begin{tabular}{lcc}
        \toprule \toprule
        \textbf{Goal $z$ (m)} & \textbf{Non-Morphing Cost (J)} & \textbf{Morphing Cost (J)} \\
        \midrule
        0.5 & 0.1338 & 0.1338 \\
        0.7 & 0.3144 & 0.3144 \\
        0.9 & Infeasible & 8.0001 \\
        \bottomrule \bottomrule
    \end{tabular}
\end{table}

\subsection{Obstacle Avoidance}

The obstacle avoidance task serves as a high-level demonstration of dynamic maneuverability of morphing aircraft, integrating the constraints and capabilities identified in the previous analyses. The task aims to navigate from a specified start position to a goal state while maintaining a collision-free path. 

\paragraph{Problem Formulation}
The trajectory optimization problem as defined in Eq.~\eqref{eqn_to_implicit}, is formulated with the cost function $J$ that represents the integral of the transient cost $L(\mathbf{x}, \mathbf{u})$, defined here as the control work. A terminal state constraint $\mathbf{x}(T) = \mathbf{x}_{\text{goal}}$ is applied for the goal position, and a path constraint $\mathbf{x}(t) \in \mathcal{X} \cap \mathcal{O}^c$ ensures obstacle avoidance. Here, $\mathcal{X}$ denotes the general physical constraints and $\mathcal{O}$ represents the obstacle region. 

To account for the vehicle's physical dimensions, the collision constraint incorporates the wing semi-span $\frac{b}{2}$, while treating the center of mass as the reference point. The obstacle is modeled as a cylinder with an effective radius $R_{\text{eff}} = R_{\text{obs}} + \frac{b}{2}$ and height $h_\text{obs}$, centered at $(x_{\text{obs}}, y_{\text{obs}}, z_{\text{obs}})$. The obstacle region is defined as:

$$ 
\mathcal{O} = \left\{ \mathbf{x} \ \middle| \ \sqrt{(x - x_{\text{obs}})^2 + (y - y_{\text{obs}})^2} \leq R_{\text{eff}}, \ z_{\text{obs}} - \frac{h_{\text{obs}}}{2} \leq z \leq z_{\text{obs}} + \frac{h_{\text{obs}}}{2} \right\} 
$$

The terminal constraint is defined as a circular region in the $y$-$z$ plane, centered at the specified coordinates $(y_{\text{goal}}, z_{\text{goal}})$ with an allowable tolerance radius of $\epsilon = 0.15$ m. However, the longitudinal position $x$ is not constrained to a fixed terminal value, as thrust is not used as a control input. Instead, a terminal region is defined to ensure the aircraft successfully navigates past the obstacle:
$$
\sqrt{(y(T) - y_{\text{goal}})^2 + (z(T) - z_{\text{goal}})^2} \leq \epsilon, \quad x(T) \ge x_{\text{obs}}
$$

The cylinder height was set to $h_{\text{obs}}=4$~m to prevent the aircraft from passing beneath or above the obstacle, forcing a lateral maneuver around the cylinder. To ensure problem feasibility, the time of flight was extended to $2.5$~s. 

\begin{figure}[hbt!]
    \centering
    \includegraphics[width=0.9\linewidth]{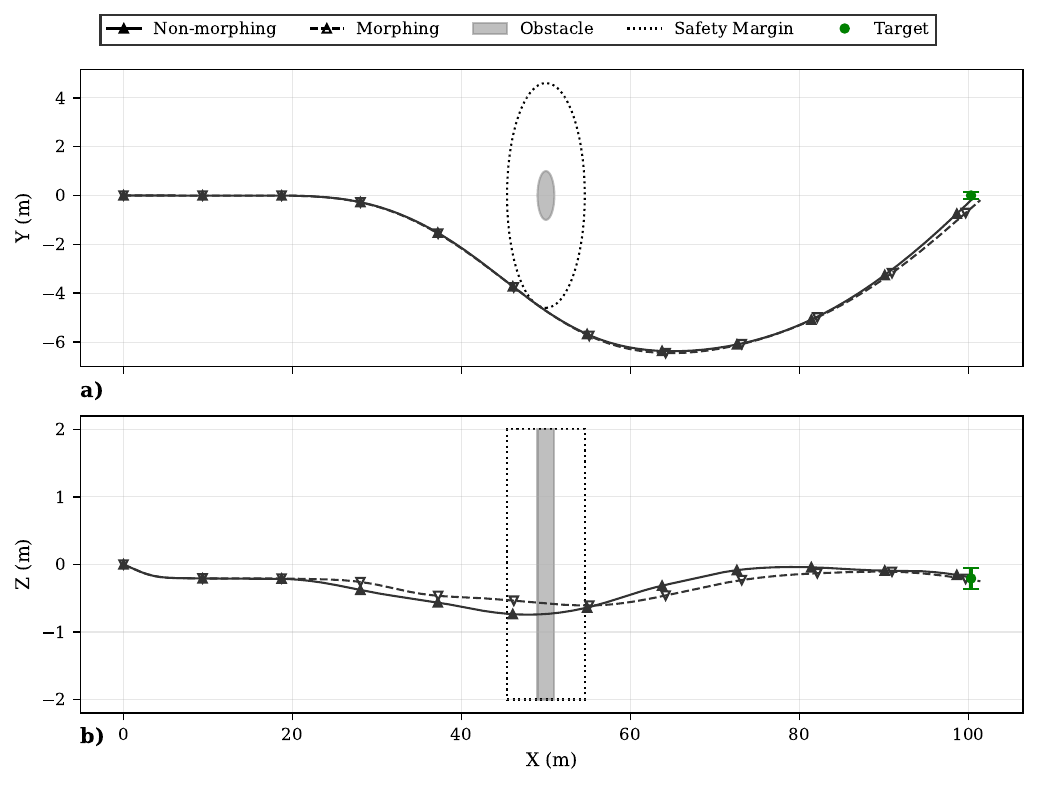}
    \caption{Trajectory over time for obstacle avoidance: (a) top-view, (b) side-view.}
    \label{fig:obs_avoidance_2/trajectory_2d_views}
\end{figure}

\begin{figure}[hbt!]
    \centering
    \includegraphics[width=\linewidth]{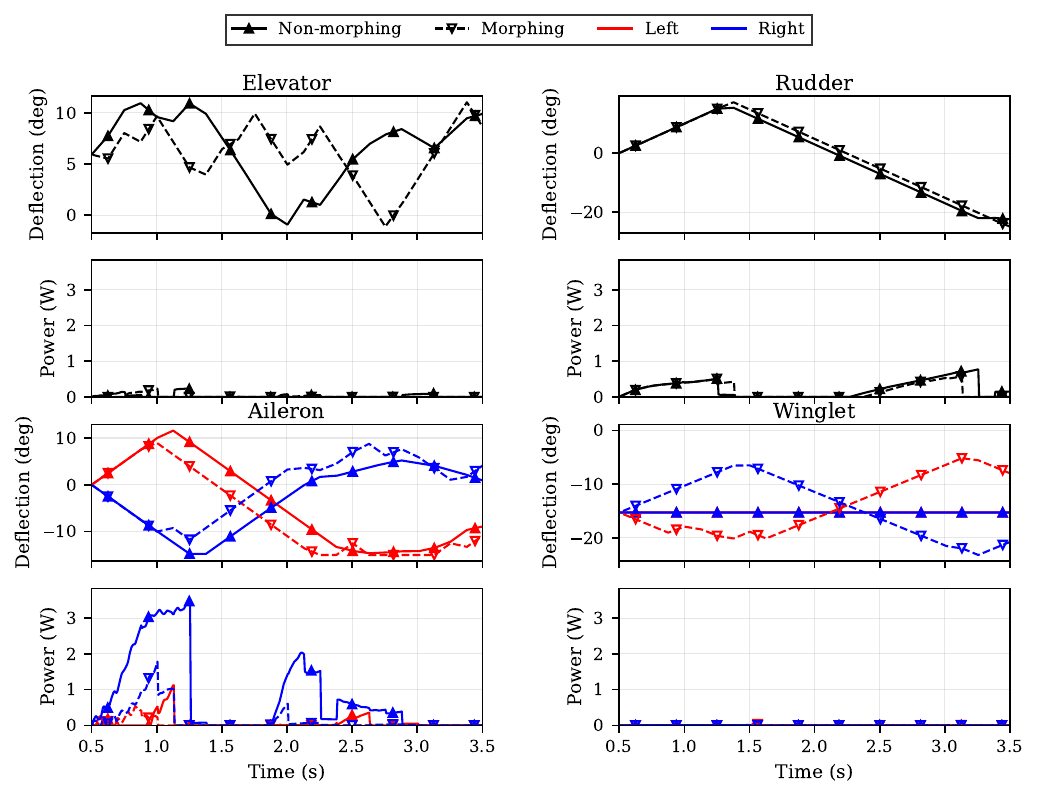}
    \caption{Optimized controls and corresponding power consumption over time for obstacle avoidance.}
    \label{fig:obs_avoidance_2/all_controls_combined_surfaces}
\end{figure}

\begin{figure}[hbt!]
    \centering
    \includegraphics[width=0.5\linewidth]{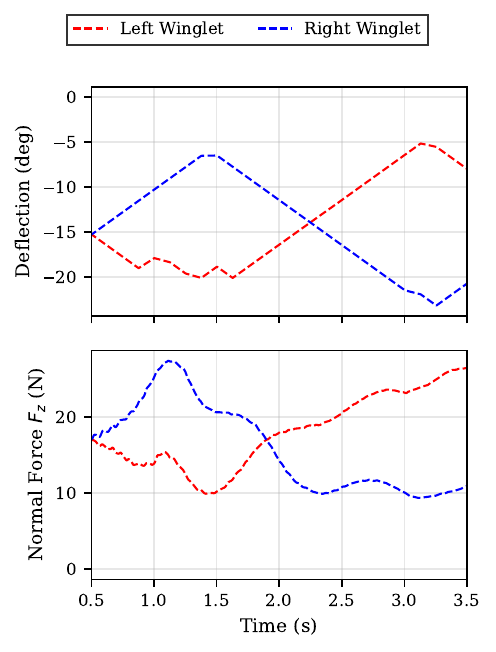}
    \caption{Winglet deflections and normal forces for morphing configuration in obstacle avoidance.}
    \label{fig:obs_avoidance_2/winglet_force}
\end{figure}

\paragraph{Performance and Cost Analysis} 
The optimized trajectories for both configurations are illustrated in Fig.~\ref{fig:obs_avoidance_2/trajectory_2d_views}, with the corresponding optimized controls and their power consumptions shown in Fig.~\ref{fig:obs_avoidance_2/all_controls_combined_surfaces}. Here, the advantages of the morphing configuration become most distinct. The aircraft exploits the inherent aero-mechanical coupling between the ailerons and the morphing winglets, effectively minimizing control cost while increasing maneuverability. For instance, by deflecting the ailerons in opposition of the winglets throughout the trajectory, the control effort of the winglets is essentially nullified; the aerodynamic load has been alleviated so much that the aircraft can morph its winglets without any power consumption. Furthermore, as seen in Fig.~\ref{fig:obs_avoidance_2/winglet_force}, the simultaneous winglet deflections also manipulate the lift forces encountered by the ailerons, assisting their motion and lowering the cost of operation. This coordination allows the morphing aircraft to leverage the powerful morphing winglets to maneuver, while still being more cost-efficient compared to the non-morphing case -- demonstrating the full capabilities of a morphing configuration. 

\begin{figure}[hbt!]
    \centering
    \includegraphics[width=0.5\linewidth]{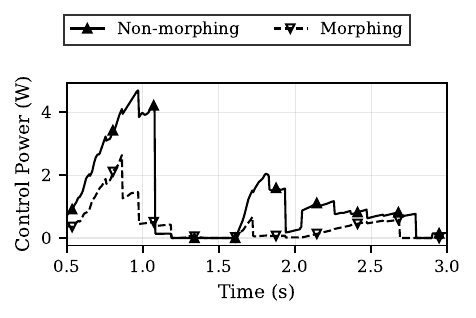}
    \caption{Total power consumption over time for obstacle avoidance.}
    \label{fig:obs_avoidance_2/control_power_vs_time}
\end{figure}

Consequently, the morphing configuration achieves a substantial 65.65\% reduction in total control cost, requiring only 1.1774~J compared to 3.4279~J for the non-morphing case. The efficiency gain is further evidenced by the total power consumption profile shown in Figure~\ref{fig:obs_avoidance_2/control_power_vs_time}, where the power required for the morphing aircraft is consistently lower than or equal to that of the non-morphing baseline throughout the entire trajectory. This reduction was achieved through the integrated trajectory optimization process and is otherwise unattainable.

\section{Conclusion}

In this study, trajectory optimization was employed to investigate the maneuverability of a morphing aircraft with active winglet dihedral morphing. We integrated an open-source aeroservoelastic model based on multi-body dynamics with a trajectory optimization framework and a detailed control cost model that captures actuation efforts. This approach allowed for the assessment of the morphing aircraft in various scenarios, including typical maneuvers, reachability analysis, and obstacle avoidance. The results of these assessments support several key conclusions regarding the efficacy and trade-offs of the morphing configuration.

First, the trim analysis demonstrates that morphing winglets fundamentally alter the aircraft's steady-state capabilities. While the fixed-wing configurations were constrained to single trim speeds due to the rigid coupling between lift and pitch dynamics, the morphing configuration successfully achieved dynamic trim across a broad velocity range ($27.5$ to $40$~m/s). By actively manipulating the spanwise lift distribution, the winglets decouple the lift requirements from the elevator, allowing the tail to focus exclusively on pitch stability. This effectively expands the operational flight envelope, particularly in high-speed regimes where aeroelastic effects would otherwise deteriorate trim.

Second, the analysis of typical maneuvers revealed distinct characteristics regarding the trade-off between performance augmentation and control cost. In the pull-up maneuver, the morphing configuration achieved a $28.95\%$ greater altitude gain than the fixed-wing baseline. However, this performance enhancement comes at a high cost; the morphing configuration incurred approximately 40 times more control cost than the conventional aircraft to execute the maneuver. Conversely, the banked turn maneuver demonstrated that morphing can yield efficiency gains alongside performance improvements. Unlike the pull-up case, the morphing configuration achieved an $8.62\%$ increase in lateral displacement while simultaneously reducing the total control cost by $13.40\%$. This suggests that for certain maneuvers, the dynamic benefits of the winglets outweigh the actuation penalty.

Third, the reachability analysis for the pull-up maneuver demonstrated that while morphing enables highly demanding maneuvers ($z_{\text{goal}} = 0.9$~m) that are physically unreachable for the fixed-wing baseline at the expense of additional control cost, it also provides the operational flexibility to remain inactive during less demanding tasks ($z_{\text{goal}} = 0.5$~m and $0.7$~m).

Fourth, the obstacle avoidance and cost analyses reveal a non-trivial benefit: morphing can function as an energy-saving mechanism in complex scenarios. While individual actuation is expensive, the obstacle avoidance study showed that variable dihedral reduced total control cost by $65.65\%$ during lateral navigation. By offloading aerodynamic saturation from the ailerons, the morphing winglets allowed for a more efficient distribution of control effort. Furthermore, the fundamental power analysis identified that exploiting the aerodynamic coupling between the aileron and winglet, specifically through opposing deflections, can neutralize hinge moments, effectively reducing control cost.

In summary, this work establishes that active morphing winglets fundamentally expand the operational flight envelope by decoupling lift and pitch requirements, thereby enhancing trim capability across a broader speed regime. Beyond expanding the steady-state envelope, morphing increases maneuverability and reachability, enabling the execution of trajectories that are infeasible for fixed-wing aircraft. Crucially, the control cost penalty associated with morphing is not absolute; as demonstrated in the turn and lateral obstacle avoidance scenarios, strategically exploiting the aerodynamic coupling between surfaces to offload hinge moments can drastically reduce control cost. This demonstrates that with intelligent control allocation, high-authority morphing can be both operationally superior and energetically efficient.

More importantly, this study highlights the necessity of combining trajectory optimization and aeroservoelastic model in the quantification of dynamic maneuver capabilities of morphing aircraft; this suggests that the same combination should be involved for the multi-disciplinary design optimization of such vehicles.  However, an apparent computational roadblock is the high computational cost of high-fidelity trajectory optimization. Such roadblock is likely best tackled by reduced-order modeling (ROM) methods, such as Ref.~\cite{vargas2023physics}, to produce an accurate and computationally efficient aeroservoelastic model that is generalizable to design parameters and hence amenable for design optimization.

\section*{Appendix A: Formulation of the Trim Cost Function}

The objective function \( J_{\text{trim}} \) is designed to quantify the ``steadiness'' of the aircraft trajectory over the post-transient interval \( \mathcal{T} = [t_s, t_f] \). It evaluates two primary characteristics for the altitude \( z \) and pitch angle \( \theta \): the long-term drift (trend) and the high-frequency unsteadiness (oscillation).

To ensure consistent weighting between state variables with different physical units, the responses are first normalized by their maximum absolute amplitude over the interval:
\begin{equation}
    \hat{\chi}(t) = \frac{\chi(t)}{\max_{\tau \in \mathcal{T}} |\chi(\tau)| + \epsilon}, \quad \text{for } \chi \in \{z, \theta\}
\end{equation}
where \( \epsilon \) is a regularization constant.

\paragraph{Linear Drift Decomposition}
We approximate the trajectory of each normalized state \( \hat{\chi}(t) \) using a linear model representing the steady-state value and a constant drift rate:
\begin{equation}
    \hat{\chi}(t) \approx \hat{\chi}_0 + \dot{\hat{\chi}}_{\text{drift}} \, t
\end{equation}
Here, \( \hat{\chi}_0 \) represents the effective steady-state intercept, and \( \dot{\hat{\chi}}_{\text{drift}} \) represents the linear drift velocity. These parameters are determined by solving the least-squares minimization problem over the time horizon:
\begin{equation}
    \{ \hat{\chi}_0, \dot{\hat{\chi}}_{\text{drift}} \} = \operatorname*{arg\,min}_{a, b} \int_{t_s}^{t_f} \left( \hat{\chi}(t) - (a + b t) \right)^2 dt
\end{equation}
A perfect trim condition implies that the drift velocity \( \dot{\hat{\chi}}_{\text{drift}} \) is zero.

\paragraph{Oscillation Metric}
Residual unsteadiness that is not captured by the linear trend is quantified using the variance of the signal:
\begin{equation}
    \sigma_{\hat{\chi}}^2 = \frac{1}{t_f - t_s} \int_{t_s}^{t_f} \left( \hat{\chi}(t) - \mu_{\hat{\chi}} \right)^2 dt
\end{equation}
where \( \mu_{\hat{\chi}} \) is the temporal mean of the signal. A perfect trim condition implies that the oscillation metric $\sigma_{\hat{\chi}}^2$ is zero.

\paragraph{Total Cost}
The total trim cost is the weighted sum of the squared drift velocity and the variance for both altitude and pitch:
\begin{equation}
    J_{\text{trim}} = \sum_{\chi \in \{z, \theta\}} \left( w_{\chi}^{\text{trend}} (\dot{\hat{\chi}}_{\text{drift}})^2 + w_{\chi}^{\text{osc}} \sigma_{\hat{\chi}}^2 \right)
\end{equation}
where $w_{\chi}^{\text{trend}}$ and $w_{\chi}^{\text{osc}}$ are the weighting factors for the drift (trend) and oscillation (variance) components of the state variable $\chi$, respectively. The values used for the regularization constant and the weighting factors in the current study are $\epsilon = 10^{-6}$, $w_{z}^{\text{trend}} = w_{\theta}^{\text{trend}} = 2.0$, and $w_{z}^{\text{osc}} = w_{\theta}^{\text{osc}} = 1.0$.

\section*{Acknowledgments}

This work is supported by NSF CMMI-2340266.  The authors also acknowledge the computing resources provided by the Institute for Computational and Data Sciences at Penn State.

\bibliography{references}

\end{document}